\documentstyle[12pt,astron]{article}
\newcommand{\beq}{\begin{equation}}
\newcommand{\eeq}{\end{equation}}
\newcommand{\bdis}{\begin{displaymath}}
\newcommand{\edis}{\end{displaymath}}
\newcommand{\bea}{\begin{eqnarray}}
\newcommand{\eea}{\end{eqnarray}}
\newcommand{\barr}{\begin{array}}
\newcommand{\earr}{\end{array}}
\newcommand{\beas}{\begin{eqnarray*}}
\newcommand{\eeas}{\end{eqnarray*}}
\newcommand{\bes}{\begin{subequations}}
\newcommand{\ees}{\end{subequations}}

\begin{document}

\begin{center}
{\large \bf NONLINEAR DYNAMICS OF MOVING CURVES AND SURFACES:
APPLICATION TO PHYSICAL SYSTEMS}\\
\vskip 1cm 
S. MURUGESH and M. LAKSHMANAN\\
{\it Centre for Nonlinear Dynamics, Department of Physics,} \\
{\it Bharathidasan University, Tiruchirappalli 620 024, India}\\
{\it murugesh,lakshman@cnld.bdu.ac.in}
\end{center}

\begin{abstract}
The subject of moving curves (and surfaces) in three dimensional space (3-D) is
a fascinating topic not only because it represents typical nonlinear dynamical
systems in classical mechanics, but also finds important applications in a 
variety of physical problems in different disciplines. Making use of the 
underlying geometry, one can very often relate the associated evolution 
equations to many interesting nonlinear evolution equations, including
soliton possessing nonlinear dynamical systems. Typical examples
include dynamics of filament vortices in ordinary and superfluids, spin 
systems, phases in classical optics, various systems encountered in physics of
soft matter, etc. Such interrelations between 
geometric evolution and physical systems have yielded considerable insight into
the underlying dynamics. We present a succinct tutorial analysis of these
developments in this article, and indicate further directions. We also point 
out how evolution equations for moving surfaces are often intimately related
to soliton equations in higher dimensions. 
\end{abstract}

\flushleft{Running Title:~~{\it Nonlinear Dynamics of moving curves and 
surfaces}}
\newpage
\section{Introduction}

Curves in 3-D correspond to one of the simplest set of objects imaginable in 
geometry. 
Commonly encountered physical systems which can be modeled on curves vary over 
a spectrum of subjects ranging from physical, chemical to life sciences. 
Examples include vortex filaments in ordinary and superfluids 
\cite{has:1972,sch:1998}, scroll waves over fluid surfaces \cite{cro:1993}, 
polymers \cite{bird:1997} and biological macromolecules such as proteins and 
DNA. 
A not too obvious example is the classical version of Heisenberg 
ferromagnetic spin lattice in its continuum limit \cite{ml:1976}, which will 
be discussed in detail later. 
All these above stated examples pose challenging problems in physics with 
regard to their dynamics due to their highly nonlinear character. Besides
these examples, the curve as an abstract geometrical object poses several 
interesting questions associated with its geometry for the mathematically 
inclined. The subject has held the attention of physicists, applied 
mathematicians and
the nonlinear dynamics community in particular, for, as we shall see, the 
equations governing their dynamics are inherently nonlinear. In certain 
physical systems the models closely resemble soliton possessing 
{\it integrable}
systems. These models prove to be a useful starting point in unraveling their
apparently complex behavior. Further, the ideas developed for curves
provide a good lead to their generalization and extension to higher dimensional
geometric objects such as surfaces and hyper-surfaces. 

In this article, after a brief description of the geometry of moving curves, 
we review some of 
the interesting intrinsic quantities associated with them that are 
frequently encountered, followed by a pedagogical analysis of several 
applications to physical systems wherein substantial results have been 
obtained. We also indicate possible generalizations to surfaces. It is the aim 
of this article to point out that it is possible to study the dynamics of 
several physical systems with inherent nonlinearity at a fundamental 
level by identifying their underlying geometry. 

The paper is organized as follows. In section \ref{sec2} we discuss the basic 
geometry of a curve in 3-D and introduce the Frenet-Serret equations which 
describe it.
In section \ref{sec3} we further discuss a few (global) geometrical quantities
that are frequently encountered. Connection to moving curves in fluid 
mechanics has yielded significant results in the past. This will be
discussed in section \ref{fluid}. Section \ref{spin} discusses the dynamics of
the 1-D classical Heisenberg ferromagnetic spin chain in its continuum limit 
which has been shown to be integrable through its connection with curve motion.
Polarized beams propagating along a curved path acquire a geometric
phase depending on the geometry of the path taken. This phase is discussed
in section \ref{opt}. Recently there has been several results in the physics
of soft matter that can be modeled on curve motion, which will be discussed
in section \ref{soft}. Section \ref{surf} discusses the extension of these
ideas to the geometry of surfaces. We conclude in section \ref{conc} with a 
summary and discussion.

\section{Geometry of Moving Curves and Soliton Equations}\label{sec2}
Consider a curve in 3-D represented by the position vector ${\bf R(\theta)}$, 
where $\theta$ is some parameter along the length of the curve. It is possible 
in principle to switch over to the arc length parameter $s$, such that the 
tangent vector ${\bf e}_1(s)$ defined as (see Fig. 1)
\beq
{\bf e}_1(s) \equiv \frac{d{\bf R}}{d s}
\eeq
is of unit length all along the curve. With this definition of the unit
tangent vector one can canonically define a unit normal vector ${\bf e}_2(s)$, 
orthogonal to ${\bf e}_1$, such that
\beq\label{fs1}
{\bf e}_{1s} = \kappa{\bf e}_2.
\eeq
Here (and henceforth), the subscript denotes derivative with respect to 
$s$. Notice that the derivative of ${\bf e}_1$ does not have a component
along ${\bf e}_1$ itself owing to its constant modulus. This follows for the
unit normal ${\bf e}_2$ too, and also for the unit binormal ${\bf e}_3$ to be 
defined 
next. The quantity $\kappa(s) = |{\bf e}_{1s}|$, appearing in Eq. (\ref{fs1}),
is the {\it curvature} (or the inverse radius of curvature) at a point $s$,
measuring the deviation of the curve from linearity. With these definitions for
the unit tangent and unit normal vectors
one can complete an ordered triad of orthonormal vectors at any point on the 
curve with the introduction of the third unit binormal vector
${\bf e}_3\equiv{\bf e}_1\times{\bf e}_2$. The derivatives of the unit normal 
and binormal can be written as
\beq\label{fs2}
{\bf e}_{2s} = -\kappa{\bf e}_1 + \tau{\bf e}_3
\eeq
\beq\label{fs3}
{\bf e}_{3s} = -\tau{\bf e}_2.
\eeq
Eq. (\ref{fs2}) introduces {\it torsion},
\mbox{$\tau={\bf e}_1.{\bf e}_{1s}\times{\bf e}_{1ss}/|{\bf e}_{1s}|^2$,} 
while 
Eq. (\ref{fs3}) follows immediately on account of orthonormality. Just as 
curvature measures deviation of a curve from linearity, torsion measures 
deviation from planarity. A planar curve has zero torsion. 

\noindent
Eqs. (\ref{fs1}-\ref{fs3}), constitute the so called Frenet-Serret equations 
(FSE) describing the basic geometry of a
curve in 3-D \cite{stok:1969,str:1961}. $\kappa$ and $\tau$ are intrinsic 
quantities describing the curve. Though the three unit vectors can be 
canonically defined at any point, 
ambiguity arises at points where $\kappa$ vanishes. Thus for a straight line,
or along segments of a curve that are straight, 
one does not have a unique choice of ${\bf e}_2$ and ${\bf e}_3$. We shall 
assume hereonwards that $\kappa$ does not vanish and is positive. 
Given the two intrinsic parameters $\kappa$ and $\tau$, at least in principle, 
a corresponding curve can be constructed, while the actual construction can be 
much involved. This {\it reconstruction problem} remains a challenge, except
for some simple curves, and has deeper significance in soliton theory 
\cite{sym:1985}. Besides the abstract interest in the subject, the problem
is also of interest in the image processing community \cite{sap:2001}. 

In the same spirit one can also consider describing the time evolution of the 
curve. First we note that the FSE Eqs. (\ref{fs1}-\ref{fs3}) can be written 
compactly as 
\beq\label{fsc}
{\bf e}_{is} = \vec{\Omega}\times{\bf e}_i~;~~i=1,2,3.
\eeq
where the Darboux vector $\vec{\Omega}=\kappa{\bf e}_3+\tau{\bf e}_1$, or simply 
$(\tau,0,\kappa)$ in the ${\bf e}_i$ basis. Let the dynamics of the curve 
${\bf R}(s)$ be given by
\beq
{\bf R}_t = \alpha{\bf e}_1 + \beta{\bf e}_2 + \gamma{\bf e}_3,
\eeq
where now the subscript $t$ stands for time derivative and $\alpha, \beta$ and
$\gamma$ are arbitrary functions of $s$ and $t$. If one assumes the
curve to be non-stretching, then $s$, and hence $|{\bf e}_i|$, are independent 
of time. Though this appears to be a serious restriction at first, a 
variety of systems still can be modeled by such curves, while for most other 
systems this turns out to be a good approximation, and definitely a good 
starting point to familiarize with the dynamics of curves. 

The dynamics of the triad can be described by defining a new vector 
$\vec{\omega} =(\omega_1,\omega_2,\omega_3)$ such that, similar to Eq. 
(\ref{fsc}), 
\beq\label{fsd}
{\bf e}_{it} = \vec{\Omega}\times{\bf e}_i~;~~i=1,2,3.
\eeq
Note that Eq. (\ref{fsd}) can be identified with the equations of a rotating 
rigid body, whose dynamics
is described by the three principal axes. The vector $\vec{\omega}$ then 
plays the
role of its angular velocity. Consequently, a moving curve can be identified
with a rotating rigid body moving along the curve \cite{ml:1979}. 

The compatibility condition $({\bf e}_i)_{st}=({\bf e}_i)_{ts}, i=1,2,3,$ then 
yields the following equation satisfied by $\vec{\Omega}$ and $\vec{\omega}$. 
\beq\label{cons}
\vec{\Omega}_t - \vec{\omega}_s = \vec{\omega}\times\vec{\Omega}.
\eeq
These are three equations for the five quantities $\vec{\Omega}$ and $\vec{\omega}$. 
Written explicitly, Eq. (\ref{cons}) reads
\bea\label{cons12}
\kappa_t - \omega_{3s} = -\tau\omega_2;\\
~~\tau_t - \omega_{1s} = \kappa\omega_2,\\
\omega_{2s} = \kappa\omega_1 - \tau\omega_3
\eea
Any 
nontrivial choice for $\omega_i$'s in terms of $\kappa$ and $\tau$ then leads,
through Eq. (\ref{cons}), to nonlinear time evolution equations for $\kappa$ 
and $\tau$ - the intrinsic variables representing a curve. Notice that 
nonlinearity is inherent in the system due to the vector product on the right 
hand side in Eq. (\ref{cons}). The compatibility conditions
on ${\bf R}$, (${\bf R}_{st}={\bf R}_{ts}, {\bf R}_{sst}={\bf R}_{tss}, etc.,$)
lead to equations relating $\alpha, \beta$ and $\gamma$ with $\vec{\Omega}$ and
$\vec{\omega}$, which however we shall not be making much use of here. 
${\bf R}(s,t)$ can
also be seen as a surface in 3-D generated by a curve. Evidently this surface
is special, being generated by a curve and one that is non-stretching. 

We list below some important examples of ubiquitous integrable nonlinear 
evolution equations admitting soliton solutions \cite{abl:1992,raj:2003} 
that can be identified with moving curves, for suitable choices of 
$\vec{\omega}$ as functions of $\kappa$ and $\tau$ 
\cite{ml:1981,lamb:1977}.

\noindent
(i) {\it Nonlinear Schr\"{o}dinger equation} (NLS)
\cite{has:1972,ml1:1977,ml:1981} :
For the choice 
\beq
\omega_1 = \frac{\kappa_{ss}}{\kappa} - \tau^2~;~~\omega_2 = -\kappa_s~;
~~\omega_3 = -\kappa\tau
\eeq
Eqs. (\ref{cons12}) lead to the NLS equation 
\beq
i\psi_t + \psi_{ss} + \frac{1}{2}|\psi|^2\psi = 0~;~~\psi 
=\kappa\exp{i\int^s_{-\infty}\tau~ds'}.
\eeq
(ii) {\it Modified Korteweg-de Vries equation} (MKdV)\cite{seg:1992,ml:1981}:
The modified Korteweg-de Vries  
\beq
u_t + 6u^2u_s + u_{sss} = 0~;
\eeq
is obtained for the choice
\beq
\omega_1 = -\frac{\tau\kappa^2}{2}~;~~\omega_2 = -\kappa_s\tau~;
~~\omega_3 = -\kappa_{ss} - \frac{\kappa^2}{2} + \tau^2\kappa,
\eeq
with $\kappa=2u$ and $\tau=constant$.\\
(iii) {\it sine-Gordon equation} \cite{lamb:1977,ml:1981}:
The sine-Gordon equation 
\beq
\theta_{st} = \sin\theta~;~~
\eeq
is obtained with the choice
\beq
\omega_1 = -\frac{1}{\tau}\cos u ~;~~\omega_2 = \frac{1}{\tau}\sin u~;
~~\omega_3 = 0
\eeq
with $\kappa=u_s$ and $\tau=constant$.

Besides these, many other evolution equations \cite{vij:1998} and certain 
integrable hierarchies have been shown to be related to curve dynamics 
\cite{seg:1992,gold:1991}.
As we noted above, the nonlinearity in the equations is due to the vector 
product in Eq. (\ref{cons}). They continue to be nonlinear even on relaxing the
condition of local length preservation. In considering applications to certain 
physical systems this becomes inevitable.

\section{Intrinsic Global Quantities Associated with Curves}\label{sec3}
In this section we introduce a few global quantities, associated with the 
geometry of the curve, which one often encounters and which have important
physical relevance in applications. The first of these is
the integral of curvature, the total curvature
\beq\label{int1}
N = \int^L_0{\kappa(s)~ds}.
\eeq
The geometrical meaning of this integral is more apparent from expressing
the integrand in terms of the tangent vector ${\bf e}_1$, namely 
$|d{\bf e}_1|=\kappa~ds$. Since ${\bf e}_1$ is a unit vector in 3-D, it can be 
seen as living on a unit sphere $S^2$. For a given curve ${\bf R}(s)$ then, the
tangent vector traces a curve, the {\it tangent indicatrix},  on the surface of
this unit sphere \cite{stok:1969}. The integrand
in Eq. (\ref{int1}) is indeed the length of this curve. For a closed curve in 
3-D we have a closed indicatrix on the unit sphere (see Fig. 2), and
\beq
N = \oint{\kappa(s)~ds}.
\eeq
However, it should be remembered that the converse is not always true, i.e., a 
closed tangent
indicatrix does not always correspond to a closed curve in 3-D. 
Through a simple argument it can be shown that, for a closed curve in 3-D, 
the total curvature $N\ge 2\pi$. This is the Frenchel's theorem, which is due
to the fact that for a closed curve in 3-D, the corresponding curve on the 
unit sphere cannot entirely lie in one hemisphere \cite{kam:2002}. Thus the 
total curvature (or the length of the closed circle traced out on the unit 
sphere) has to be greater that the circumference of the great circle which is 
indeed $2\pi$. 


The second integral we shall consider, and one that is frequently encountered, 
is the 'energy' integral
\beq\label{int2}
E = \int{\kappa^2~ds} = \int{\big[\frac{d{\bf e}_1}{ds}\big]^2~ds}.
\eeq
We call this the energy integral for this is the first relevant term for 
any typical physical system modeled by a curve. This is understandable, since
any change in shape of the curve requires energy, whose first nontrivial term
has to be quadratic. The simplest example is an elastic band, for which this is
just the potential due to the Hooke's law. Further, if one imagines $s$ to 
represent time and ${\bf e}_1$ as the position vector of a particle, then the 
integrand in Eq. (\ref{int2}) is just its kinetic energy (apart from a factor)
and $E$ is the action. Since ${\bf e}_1$ moves on a unit sphere, this is just 
the action integral of a particle constrained to move 
on the surface of a sphere. $E$ turns out to be the Hamiltonian for the 
classical continuous Heisenberg model for spin systems in 1-D, of which more 
will be said in section \ref{spin}. 

It may be naturally tempting at this stage to see the geometrical relevance of
the other integral, namely that of torsion 
$\tau$
\beq\label{ph}
M = \int^L_0{\tau~ds}. 
\eeq
Analogous to Eq. (\ref{int1}), this can be seen as the length of the 
{\it binormal indicatrix}. However, $M$ has a better geometrical interpretation
which shows up in several practical situations. 
The answer can be found in Eq. (\ref{fsc}) and noting that $\tau$ is the 
component of $\vec{\Omega}$ along the tangent direction. Thus torsion is the local
phase acquired by the normal and the binormal as we move along the curve. Eq. 
(\ref{ph}) gives the total phase acquired by the $({\bf e}_1-{\bf e}_2)$ plane
as we move through the total length of the curve, compared to the same plane in
a non-rotating basis. Further, parameterizing the tangent vector in terms of 
spherical polar coordinates, (i.e., 
${\bf e}_1=(\sin\theta\cos\phi,\sin\theta\sin\phi,\cos\theta)$), we find for
a closed tangent indicatrix
\beq
M = \oint{\tau~ds} = \oint{\cos\theta~d\phi}.
\eeq
Using Stoke's theorem, the integral above can be rewritten as
\beq
M = \oint{\cos\theta~d\phi} = 2\pi - \int_S{\sin\theta d\theta d\phi}.
\eeq
The last integral in the above equation is just the area enclosed by the 
tangent indicatrix on the unit sphere. Thus, except for an additive $2\pi$, 
$M$ gives the area enclosed by a closed tangent indicatrix. We shall encounter 
the same in section \ref{opt} when we discuss applications to geometric phases
in classical optics. In the next few sections we discuss some physical systems 
that can be modeled using space curves, where significant results have been 
reported. 

\section{Vortex Filaments}\label{fluid}
Dynamics of a fluid is described by its velocity field ${\bf V}$. Vorticity is
defined as the curl of velocity ${\bf W} (=\nabla\times{\bf V})$. At low 
velocities 
vortices are generally observed when a fluid is subject to rotation, the most 
common example being the kitchen sink vortex, while at high velocities, 
rotation is inherent in the fluid. They are known to occur in various forms 
such as lines, rings, etc., \cite{batch:1967}. In the following we shall 
confine ourselves only to line vortices.
Vortices in ordinary fluids have fascinated thinkers over several centuries
dating back to the masters in Rome. Some early detailed observations on fluid
vortices can be found in Learnado da Vinci's writings. Besides their common
occurrence, the geometrical features associated are shared by several other 
systems depicting vortex behavior. These include large scale systems such
as twisters, magnetic flux tubes in solar corona \cite{spr:1981}, systems at
microscopic scale such as Bose-Einstein condensates\cite{dali:2003},
superfluid helium \cite{sch:1998} and magnetic media. Besides the intrinsic
interest in them
understanding their dynamics has been realized to be the key to fluid 
turbulence  \cite{fri:1996}.

As is well known, motion of incompressible fluids is described by the Euler 
equations
\cite{saff:1992}. 
For such a fluid $\nabla.{\bf V}=0$, then it follows from the
Helmholtz theorem that the velocity field at a point ${\bf X}$ in the fluid 
is derived entirely from its curl - the vorticity ${\bf W}$,
\beq
{\bf V(X)} = \frac{1}{4\pi}
\int_V{\frac{{\bf (X-X')}\times{\bf W(X')}}{|{\bf X-X'}|^3}d^3{\bf X'}}.
\eeq
For filament vortices, the vorticity ${\bf W}$ is along the tangent direction
${\bf e}_1$, ${\bf W}=C{\bf e}_1$. 
We recall two of Helmholtz laws of vortex motion\cite{saff:1992}: i) Vortex 
lines move with the fluid. i.e., if the line vortex is described by the 
position vector ${\bf R}$, then ${\bf V(R)}={\bf R}_t$ and ii) the strength of 
the vortex tube remains constant, i.e., $C$ is a constant.
To a first approximation, known as the {\it localized induction approximation},
wherein one assumes $a)$ thinness of the core radius compared to the mean 
radius of curvature and $b)$ that the vortices are non-stretching
\cite{saff:1992}, the dynamics of these vortices are governed, with a rescaled
time, by the {\it localized induction equation}
\beq\label{lia}
{\bf R}_t = {\bf R}_s\times{\bf R}_{ss}~;~~{\bf R}_s.{\bf R}_s=1.
\eeq
Here, $s$ is 
the distance along the vortex line. The equation describes the motion of the 
vortex 
induced by its own geometry. If ${\bf R}$ is a straight line, then evidently
the dynamics is trivial. 

Mapping the vortex line on a space curve discussed in section \ref{sec2},
and interpreting ${\bf R}_s$ as the unit tangent vector ${\bf e}_1$, 
Eq. (\ref{lia}) can be identified as 
\beq\label{lie}
{\bf e}_{1t} = {\bf e}_1\times{\bf e}_{1ss}~;~~ {\bf e}_1.{\bf e}_1 =1.
\eeq
Now making use of FSE Eq. (\ref{fsc}), one can check that Eq. (\ref{lie}) can
be rewritten as 
\beq\label{e1t}
{\bf e}_{1t} = \kappa_s{\bf e}_3 - \kappa\tau{\bf e}_2
\eeq
One can complete the trihedral evolution as 
\beq
{\bf e}_{2t} = -\kappa_s{\bf e}_1 
+ (\frac{\kappa_{ss}}{\kappa} -\tau^2){\bf e}_3,
\eeq\beq
{\bf e}_{3t} = -(\frac{\kappa_{ss}}{\kappa} -\tau^2){\bf e}_2
- \kappa\tau{\bf e}_2.
\eeq
As discussed in section \ref{sec2}, the compatibility conditions 
${\bf e}_{st}={\bf e}_{ts}$ leads to the evolution equation for the curve that
can be written in terms of the intrinsic
quantities $\kappa$ and $\tau$, \cite{ml:1976}. 
\beas
\kappa_t =-(\kappa\tau)_s - \kappa_s\tau,\\
\tau_t = [(\kappa_{ss}/\kappa) - \tau^2]_s + \kappa\kappa_s.
\eeas
Further, it was shown by Hasimoto \cite{has:1972} that these equations 
can be combined to a single complex equation - the nonlinear Schr\"{o}dinger 
equation (NLS) 
\beq\label{nls}
i\psi_t + \psi_{ss} + \frac{1}{2}|\psi|^2\psi = 0,
\eeq
if one defines $\psi\equiv\kappa\exp[{\int^s_{-\infty}\tau~ds'}]$. The NLS 
equation being an integrable nonlinear evolution equation \cite{zakh:1971} 
(ie., possessing an infinite number of 
{\it independent} integrals of motion \cite{raj:2003,abl:1992}) with soliton 
solutions, similar solutions were conjectured
in line vortices also. These were subsequently observed experimentally 
\cite{hopf:1982}. The integrable nature of Eq. (\ref{lia}) also guarantees
stability of these solitonic excitations. 

The form of the vortex filament corresponding to a one-soliton solution of the
NLS is given in Fig. 3. Explicitly, the expression for the position vector
${\bf R}$ is given by
\bea\label{vortR}
{\bf R}(s,t) = \Big\{s-\mu\tanh(\nu\xi),-\mu{\rm sech}(\nu\xi)\cos\eta,
\\\nonumber
-\mu{\rm sech}(\nu\xi)\sin\eta\Big\},
\eea
where $\xi=s-2\lambda t$, $\eta=\lambda s +(\nu^2-\lambda^2)t$ and 
$\mu=2\nu/(\nu^2+\lambda^2)$, involving two constant parameters $\nu$ and 
$\lambda$.
The condition that the dynamics be length preserving is a valid approximation
for small times. Prominent examples where this condition has to be 
necessarily relaxed include superfluids and certain weakly nonlinear 
oscillations such as in chemical media \cite{sch:1998,cro:1993}. Chemical 
oscillations are governed by the complex Ginzburg-Landau equation,
\beq\label{cgle}
{A}_t = A (1+ib)\nabla A - (1+ic)|A|^2A  
\eeq
wherein numerical investigations have shown instabilities arising due to 
stretching and bending of the vortex lines \cite{aran:1993}.

Another prominent example where dynamics of vortex filaments play an 
important role is in scroll waves encountered in heart muscles. Scroll waves
known to form in the ventricle of the heart in response to certain 
stimulations (Fig. 4). Its geometry can be described by a space curve, the 
filament of the scroll, about which the scroll rotates, with a frequency
larger than the rate of heart beat. Instabilities in the scroll wave can
result in a break up into smaller independent scrolls leading to ventricular
fibrillation which often proves fatal \cite{win:1987}. 

\section{Spin Systems}\label{spin}
Magnetism is a consequence of the quantum nature of the spin angular momentum
associated with electrons. Specifically, the fundamental Hamiltonian
responsible for spin-spin interaction that leads to spontaneous magnetization
in magnetic materials, which is the Heisenberg interaction, is a consequence of
Pauli's exclusion principle \cite{kitt:1960}. If 
${\bf S}_i(=(S_i^1,S_i^2,S_i^3))$ 
were to denote the spin vector at a site denoted by $'i'$, then the isotropic
Heisenberg Hamiltonian is given by
\beq\label{hei}
H = -J\sum_{<i,j>}{\bf S}_i.{\bf S}_j~;~~|{\bf S}_i|^2=1,
\eeq
supplemented by the usual spin commutation relations 
$[S^a_i,S^b_j]=\delta_{ij}\epsilon^{abc}S^c_i$, where $\delta_{ij}$ and 
$\epsilon^{abc}$ are the usual Kronecker delta and Levi-Civita tensors,
respectively.
Here, the bracket in the sum denotes nearest neighbor. Due to the tendency of 
any system to attain a minimum energy configuration, from Eq. (\ref{hei}) we 
note that the adjacent spins tend to align parallel when $J>0$, leading to 
ferromagnetism and anti-parallel for $J<0$, giving rise to anti-ferromagnetism.
Particular cases of the 
system lead to other interesting models such as the $XY$-model, n-Potts model 
and the Ising model \cite{chai:1970}.

Simple as it may appear, the system, Eq. (\ref{hei}), has posed itself as one 
of the challenging
problems in physics. Though further complications to Eq. (\ref{hei}) may arise 
in the form of anisotropy, external magnetic field, dissipation, etc., the 
real difficulty turns out to be due to the dimension and structure of the 
lattice, besides the nonlinearity in their dynamic equations.  

From the Hamiltonian, Eq. (\ref{hei}), using the commutation relations, one
obtains the dynamical equations for the spins as 
\beq
S^a_{it} = J\sum_j\epsilon^{abc}S^b_i~S^c_j,
\eeq
where the sum runs over all nearest neighbors of the site $i$. 
Though the system is discrete and quantum in nature, for low energies it 
proves sufficient to consider the classical continuum version of the model. In
this case the dynamics of the spin field is governed by a special case of the
Landau-Lifshitz
equation (LLE) given, after appropriate scaling, by \cite{lle:1935,ml:1976}
\beq\label{lle}
{\bf S}_t = {\bf S}\times\nabla^2{\bf S}~;~~|{\bf S}|^2 = 1.
\eeq
For quasi-one dimensional magnets, the evolution equation can be written as 
\beq\label{llex}
{\bf S}_t = {\bf S}\times{\bf S}_{xx}~;~~|{\bf S}|^2 = 1. 
\eeq
Associating now a space curve with the continuum spin chain and identifying 
the spin vector ${\bf S}(x,t)$ with the unit tangent vector ${\bf e}_1$ 
and the spatial coordinate $x$ with the arc length of the curve $s$, the 
evolution equation becomes
\beq\label{lle1}
{\bf e}_{1t} = {\bf e}_1\times{\bf e}_{1ss}~;~~|{\bf e}_1|^2 = 1
\eeq

\noindent
Now making use of the earlier identification in section \ref{fluid} (see 
Eqs. (\ref{e1t} - \ref{nls})), the one 
dimensional continuum LLE is seen to be equivalent to the integrable NLS, 
which in turn implies the integrability of LLE in 1-D
\cite{ml1:1977}. 

A spin one-soliton solution is shown in Fig. 5, for which the components 
are
\bea
{\bf S} =\Big\{\mu{\rm sech}(\nu\xi)(\nu\tanh(\nu\xi)\cos\eta+\lambda\sin\eta),
\\\nonumber
\mu{\rm sech}(\nu\xi)(\nu\tanh(\nu\xi)\sin\eta-\lambda\cos\eta),\\\nonumber
1-\mu\nu{\rm sech}^2(\nu\xi)\Big\},
\eea
where $\eta, \mu$ and $\xi$ are as defined in section \ref{fluid} under 
Eq. (\ref{vortR}). 

Besides 1-D ferromagnets, curve dynamics has also been used in studying
low lying excitations in 1-D anti-ferromagnets, where it is related to a
special case of static 2-D ferromagnets \cite{radha:1993}.
Soliton solutions in 1-D spin chains are well known. However, the glossary of
results available in the literature for spin systems in higher spatial 
dimensions are hardly
substantial, though several particular cases have been effectively dealt with
\cite{kose:1990,bary:1994,vij:1998}.
Notable among these, in the present context, is that of vortex/anti-vortex 
rings in ferromagnets \cite{coop:1999}, and particular radially and 
axisymmetric solutions for the Heisenberg Hamiltonian systems \cite{kali:1979,por:1991}.

\section{Geometric Phase in Classical Optics}\label{opt}
The concept of geometric phase in quantum mechanics has gained much 
significance following the work of Berry \cite{berry:1984,shap:1989}. When a 
quantum
system is subjected to a time dependent Hamiltonian, the state of the system
acquires an extra phase, in addition to the dynamic phase, depending on the 
geometry of the time dependent parameter in the Hamiltonian. The simplest 
example is a charged particle in the presence of a time dependent magnetic 
field vector. Here, the state of the particle acquires an extra phase depending
on the geometry of the magnetic field. Though the concept has achieved much
of its significance in quantum theory, there are classical examples that do 
exhibit similar features. 

Let us consider the propagation of polarized light along a thin twisted
optical fiber (see Fig. 6).
Electromagnetic waves carry inherently three vectors that describe them, 
namely the electric field ${\bf E}$, the magnetic field ${\bf B}$ and the 
propagation vector ${\bf k}$. In free space the three vectors are orthogonal,
but acquire a component along the other vectors while traveling in a dipolar
medium. Nevertheless, one can still consider them to be orthogonal to a good
approximation, as we shall do here. If one considers the fiber to be a curve
described by the vector ${\bf R}(s)$, the direction of the propagation vector 
${\bf k}$ is simply the tangent vector ${\bf e}_1$. The ${\bf E,B}$ vectors 
acquire a phase depending on the geometry of the fiber, which has been
experimentally observed
\cite{hald:1986,berry:1987,shap:1989,kug:1988,wu:1986}. 

Unlike the normal and binormal vectors of section \ref{sec2} which were 
canonically determined purely from the geometry of the curve itself, the 
${\bf E,B}$ fields present a frame that is physically observable. Evidently, 
they can still be expressed in the Frenet-Serret basis formed by ${\bf e}_i$. 
An expression for the geometric phase acquired can be obtained from the 
wave equation. 

If ${\bf R}(s)$ were to describe the center of the fiber, then any point
on the fiber can be uniquely expressed as
\beq
{\bf X} = {\bf R}(s) + r\cos\theta{\bf e}_2 + r\sin\theta{\bf e}_3,
\eeq
where $r$ and $\theta$ are polar coordinates described on the circular 
cross-section at $s$. This holds good for fibers where the radius of curvature
at any point is much larger than the radius of the cross-section. Defining
${\bf i}_2=\cos\theta{\bf e}_2+\sin\theta{\bf e}_3$ and 
${\bf i}_3={\bf e}_1\times{\bf i}_2$, we find
\beq\label{dx}
d{\bf X} = (1-\kappa r\cos\theta)ds{\bf e}_1 + dr{\bf i}_2 + r(d\theta 
+ \tau ds){\bf i}_3.
\eeq
It is convenient to transform to orthogonal coordinates $(s,r,\phi)$, where 
$\phi = \theta +\int^s{\tau ~ds'}$. In terms of these coordinates the gradient
operator is given by
\beq\label{grad}
\nabla = \frac{{\bf e}_1}{1-\kappa r\cos\theta}\frac{d}{ds} 
+ {\bf i}_2\frac{d}{dr} + \frac{{\bf i}_3}{r}\frac{d}{d\phi}.
\eeq

\noindent
In these transformed coordinates 
the propagation of the wave, described by the wave equation for a field $\psi$,
$\nabla^2\psi=-E\psi$, is just that along a straight fiber to second order
in $\kappa$ \cite{kug:1988}. Thus the only difference in 
propagation, between a straight fiber and a curved one, is the phase factor
$\phi-\theta=\int^s{\tau~ds'}$. This is indeed the phase acquired 
by the polarization vector traveling along a curved fiber. 

Though the discussion here was
confined to plane electromagnetic waves, the result can be generalized to 
transverse acoustic modes along a curved path also, as the conclusion 
merely follows from the wave equation. Further similar geometric phases
have also been identified in a dynamic context, particularly in continuous
ferromagnetic and anti-ferromagnetic spin chains \cite{radha:1990,radha:1993}.

\section{Soft Condensed Matter}\label{soft}
The last few decades have seen a variety of mathematical methods developed and 
applied to systems that are of interest specifically in life sciences. Several 
filamental structures most common in nature, namely biological macromolecules 
such as proteins, bacterial fibers and DNA, tendrils in creeping plants and
polymers exhibit complex physical phenomena due to the high order of 
nonlinearity in their governing equations \cite{tabor:2001}. The elementary 
physical model for most of these 
systems is a thin elastic rod acted upon by forces and moments. Besides these,
as in the previous sections, we shall also assume $a)$ the cross-section to be 
thin compared to the mean radius of curvature, and $b)$ 
the rod is inextensible and unshearable. Let us identify a local basis along 
the axis of the filament $({\bf e}_1,{\bf i}_2,{\bf i}_3)$, as defined in the 
previous section. Recall that the last two vectors differ from the normal and 
binormal by an angle
$\theta$. The description of the rod dynamics is simplified by averaging out 
the equations for the force ${\bf F}$ and moment ${\bf M}$ over the 
cross-section. After appropriate scaling, the model is described by the 
Kirchoff equations \cite{ant:1995}
\bea
{\bf F}_{ss} = ({\bf e}_1)_{tt}, \\
{\bf M}_s + {\bf e_1\times F} 
= {\bf i}_1\times({\bf i}_1)_{tt} + {\bf i}_2\times({\bf i}_2)_{tt},
\eea
\beq\label{mom}
{\bf M} = a\kappa_1{\bf e}_1 + \kappa_2{\bf i}_2 + b\kappa_3{\bf i}_3.
\eeq
Here $b\equiv I_3/{I_2}$ is the ratio of the principal moments of inertia of 
the cross-section and $a\equiv J/(2I_2(1+\sigma))$, where $J$ is a factor 
depending on the particular shape of the cross section and $\sigma$ is the
Poisson ratio. For a rod of uniform circular cross section, $I_2=I_3=J/2$. 
Eq. (\ref{mom}) defines the moment ${\bf M}$ in terms of the components
of the twist vector $(\kappa_1,\kappa_2,\kappa_3)=(\tau+d\theta/{ds},
\kappa\sin\theta,\kappa\cos\theta)$. The model has been used in analyzing 
instabilities arising due to change in curvature, leading to spontaneous 
changes in its conformation. An elastic coil when stretched and untwisted 
completely and slowly released can go into various possible conformations, 
induced by variation in curvature, such as looping \cite{tab:1998b}, 
orientation reversal, buckling, etc. Typical examples include telephone coils 
and tendrils in creeping plants \cite{tab:1998}.

As we mentioned in section \ref{sec3}, the expression for $E$, 
Eq. (\ref{int2}), describes the free energy of a flexible string. A more 
realistic model involves, to quadratic order, contributions due to twist and
possible external fields, such as viscosity:
\beq
E = \int(A(\kappa-\kappa_o)^2 +C\tau^2)~ds - \int\Lambda~ds. 
\eeq
Here $A$ and $C$ represent coefficients of bend and twist respectively. 
$\kappa_o$ represents any possible ground state configuration other than a
straight rod. This is understandable in case of most polymers or proteins and
DNA, which all have a nontrivial natural configuration. The second integral 
imposes the constraint of local length preservation. The dynamic equations
for the elastica are obtained by equating the force $-\delta E/\delta {\bf R}$
to the viscous drag due to the surrounding fluid, proportional to ${\bf R}_t$. 
Note that the second derivative in time of ${\bf R}$ does not 
appear in the force equation, since the first derivative is the more dominant 
term near equilibrium. The model has been used in analyzing dynamics of 
polymers in fluids, leading to geometry induced phenomenon such as untwisting 
and folding \cite{gold:1995,gold:1998}.

\section{Extension to Surfaces}\label{surf}
The success of the models based on curve evolution prompts us to consider 
systems that can be modeled on surfaces and hyper-surfaces in higher dimension.
Indeed there has been considerable progress both in the variety of physical
systems covered in these models and on developing the required mathematical
methods to deal with these systems \cite{bish:1984,pelce:1986}.
In this section we discuss how the ideas we developed for curves can be 
extended to surfaces in 3-D. Let the surface be described by a position vector
${\bf P}(x,y)$ \cite{str:1961,stok:1969}. Just as in the case of curves, this 
allows one to define
three vectors (see Fig. 7), namely, ${\bf  {u}}\equiv \partial{\bf P}/\partial x$, 
${\bf  {v}}\equiv \partial{\bf P}/\partial y$ and a third vector orthogonal 
to these two, 
${\bf N}={\bf  {u}}\times{\bf  {v}}/|{\bf  {u}}\times{\bf  {v}}|$.
Note that unlike the case of curves, the three vectors are not orthonormal, in
general.
While ${\bf N}$ is orthogonal to both ${\bf  {u}}$ and ${\bf  {v}}$ by 
construction,
the latter two are neither orthogonal to each other nor normalized. For the 
same reason the normal vector ${\bf N}$ becomes ambiguous when ${\bf  {u}}$ 
and ${\bf  {v}}$ are parallel. We shall consider only {\it smooth} surfaces,
wherein these ambiguities do not occur. The intrinsic variables describing the 
surface are obtained from the first and second fundamental forms,
\beq
(d{\bf P})^2 = Edx^2 + 2Fdxdy + Gdy^2,
\eeq
\beq
-d{\bf P}.d{\bf N} = Ldx^2 + 2Mdxdy + Ndy^2,
\eeq
respectively.
The variation of the three vectors with respect to the two parameters is given 
by the Gauss-Weingarten equations, which again involve the coefficients of the
second fundamental form and derivatives of the first fundamental form. 
The coefficients are not completely independent and are related by the
Gauss-Codazzi-Mainardi conditions ${\bf P}_{xxy}={\bf P}_{xyx}$ and 
${\bf P}_{yyx}={\bf P}_{yxy}$.

The role of intrinsic curvature for such a surface is played by the Gaussian 
curvature, defined as 
\beq
K = \frac{LN-M^2}{EG-F^2}.
\eeq
For a flat surface, evidently  $d{\bf N}$ vanishes and the Gaussian curvature
$K$ trivially vanishes. Solitons in surface theory as specific surfaces were 
known to geometers centuries before the formal theory of soliton developed. 
Indeed if one
considers a surface of constant negative curvature (say $K=-1$) -a
pseudo spherical surface, the angle
$\theta$ between the two vectors ${\bf  {u}}$ and ${\bf  {v}}$ satisfies 
the sine-Gordon equation 
\beq
\theta_{xy} = \sin\theta.
\eeq
The sine-Gordon equation is an integrable model and its role in several 
physical systems was recognized only in the middle of the twentieth century 
\cite{chai:1970,jack:1990,raj:2003}.

The difference between a smooth surface 
and one generated by a moving curve must be noted. While a smooth surface 
can be imagined as one generated by a moving curve, the converse is not 
always true. However the surface generated by a moving non-stretching curve
has the advantage that a triad of unit vectors can be defined at every point, 
which makes it easier to deal with. Besides, the underlying curve being 
described by two intrinsic variables $\kappa$ and $\tau$, the surface can be
immediately described by a complex function such as $\psi$ in Eq. (\ref{nls}).

Consider hence a surface generated by sliding a curve of the type discussed
in section \ref{sec2} \cite{vij:1998}. Such a surface is essentially described
by a position vector with two parameters, of the type described in the previous
paragraph. However this surface is special, as at every point on the surface we
can define a triad of unit orthonormal vectors ${\bf e}_i(x,y), i=1,2,3$. If 
one considers the time evolution of this surface, these vectors are then 
described by three sets of equations
\beq\label{comp}
({\bf e}_i)_x = \vec{\Omega}\times{\bf e}_i~;~~
({\bf e}_i)_y = \vec{\Gamma}\times{\bf e}_i~;~~
({\bf e}_i)_t = \vec{\omega}\times{\bf e}_i.
\eeq
Here $\vec{\Gamma}(=(\gamma_1,\gamma_2,\gamma_3))$ plays a role similar to 
$\vec{\Omega}$ of Eq. (\ref{fsc}) and $\vec{\omega}$ of Eq. (\ref{fsd}). 
Consequently, along with Eq. (\ref{cons}), we have two more compatibility 
conditions
\beq
\vec{\omega}_y - \vec{\Gamma}_t = \vec{\Gamma}\times\vec{\omega}~;
~~\vec{\Gamma}_x - \vec{\Omega}_y = \vec{\Omega}\times\vec{\Gamma}.
\eeq
Written in terms of the intrinsic variables $\kappa$ and $\tau$, these 
evidently lead to coupled nonlinear partial differential equations in two 
spatial dimensions and one time dimension (i.e., $(2+1)$ dimensions). Several 
(2+1)-D generalizations of the Heisenberg equation
(\ref{hei}) have been mapped on such moving surfaces and further they have 
been related to integrable (2+1)-D nonlinear evolution equations by the 
above procedure. Some examples are as follows \cite{myr:1998}.

\noindent
(i) {\it Myrzakulov I equation} \cite{mir:1987}:\\
This equation reads 
\bea
{\bf S}_{t} = ({\bf S}\times{\bf S}_{y} + u{\bf S})_x,\\
u_x = -{\bf S}.{\bf S}_{x}\times{\bf S}_{y}.
\eea
Identifying again the spin vector ${\bf S}(x,y,z)$ with the unit tangent
vector ${\bf e}_1(x,y,z)$ and following a procedure similar to section 
\ref{spin}, we can make the following identifications in Eqs. (\ref {comp}).
\bea
\gamma_2 = \frac{{u}_x}{\kappa},\\
\omega_1 = \frac{\kappa_{xy}}{\kappa} - \tau\partial^{-1}_x\tau_y~;
~~\omega_2 =-\kappa_y~;~~\omega_3 = -\kappa\partial^{-1}_x\tau_y
\eea
where $\partial^{-1}_x$ stands for integration with respect to $x$. 
Analogous to the NLS equation (\ref{nls}) for the 1-D Landau-Lifshitz equation
(\ref{llex}), one can construct a complex NLS type equation - the Zakharov 
equation for a complex function $q\equiv(\kappa/2)\partial^{-1}_x(-\tau)$ 
\beq
iq_t(x,y,t) = q_{xy} + Vq~;~~V_x=2|q|^2_y.
\eeq
\noindent
(ii) {\it Ishimori II Equation}\cite{ishi:1984}:\\
The equation has the form
\bea
{\bf S}_{t} = {\bf S}\times({\bf S}_{xx} - {\bf S}_{yy})
+u_x{\bf S}_{y} + u_y{\bf S}_{x},\\
u_{xx} + u_{yy} = 2{\bf S}.{\bf S}_{x}\times{\bf S}_{y}.
\eea
The procedure adapted for (i) in this case leads to 
\bea
\gamma_2 =-\frac{u_{xx}+u_{yy}}{2\kappa}~;
~~\omega_1 = \frac{\tau\omega_3-\omega_{2x}}{\kappa}\\
\omega_2=u_x\gamma_2 -\kappa_x +\gamma_{3y}+\gamma_1\gamma_2\\
~~\omega_3=-\kappa\tau+\kappa u_y +u_x\gamma_3 -\gamma_{2y} +\gamma_2\gamma_3.
\eea
An analogous complex equation can be constructed if one defines $p=a/b$, where
\beq
a =\frac{1}{2}[\kappa^2+\gamma_2^2+\gamma_3^2-2\kappa\gamma_2]^{1/2}
\eeq
\beq
b = \partial_x^{-1}\bigg(\tau -\frac{u_y}{2}+\frac{\gamma_2\gamma_{3x}-\gamma_3\gamma_{2x}-\kappa\gamma_{3x}}{\kappa^2+\gamma_2^2+\gamma_3^2-2\kappa\gamma_2}\bigg).
\eeq
$p$ satisfies the Davey-Stewartson equation II,
\beq
ip_t + p_{xx} - p_{yy} - 2p\phi = 0,
\eeq
\beq
\phi_{xx} + \phi_{yy} + (|p|^2)_{xx} - (|p|^2)_{yy} = 0.
\eeq
Similar connection can be identified between Ishimori-I equation and the 
Davey-Stewartson equation, which are known to possess localized solutions 
called dromions. 

Though examples presented here pertained to generalized versions of spin 
systems, other physical scenarios, such as hydrodynamics, magnetohydrodynamics
and dynamics of shell membranes, have also been effectively investigated
using geometry of moving surfaces \cite{rog:2003}. 
Other alternative procedures have also been suggested to relate nonlinear
integrable evolution equations to moving surfaces 
\cite{sym:1985,naka:1993,kono:1996}. 
Notable among them is the one introduced by Konopelchenko \cite{kono:1996}.
Here, one considers surfaces induced in three dimensional space via solutions 
of two dimensional linear problems, while their integrable dynamics is obtained
by the (2+1) dimensional nonlinear evolution equations obtained as 
compatibility conditions of the linear problems. Nevertheless, the subject 
of relating surfaces to higher dimensional nonlinear partial differential
equations, and the connection to the larger problem of integrability, remains
a open and challenging task. 
 
\section{Conclusion}\label{conc}
We have attempted in this tutorial review to illustrate the subject of 
the nonlinear dynamics underlying the curve motion and discussed how this
arises as a natural model in studying various physically interesting systems.
That the evolution equations in many cases closely relate to integrable
nonlinear evolution equations with soliton solutions adds to the interest in 
the subject.
This has been illustrated by considering several systems wherein curve
geometry plays a crucial role. 
While we have been largely concerned with the simpler case of non-stretching
curves, the condition can be relaxed and generalized to include a wider 
class of systems often encountered. Such studies can also relate the dynamics
to interesting spatio-temporal patterns. The extension of curve dynamics
to surfaces as a method to 
study higher dimensional nonlinear partial differential equations, and their
integrability, has been discussed as a challenging future direction of the 
ideas well realized in the case of curves. 

\section{Acknowledgment}
The work reported here forms part of a Department of Science and Technology,
Government of India sponsored research project.

\newpage

\newpage
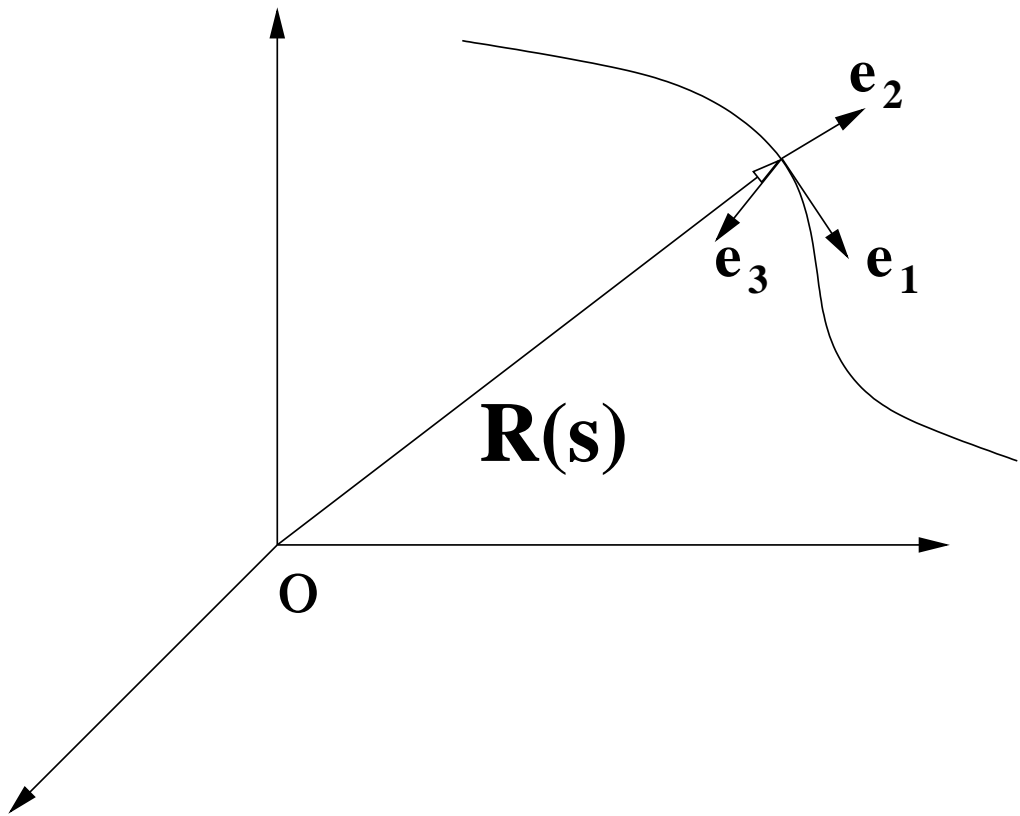
\begin{figure}
\caption{A curve in 3-D represented by the position vector 
${\bf R}$ and parametrized by the arc length $s$. An ordered triad of unit 
vectors 
${\bf e}_i, i=1,2,3$ can be canonically defined at any point on the curve.}
\end{figure}

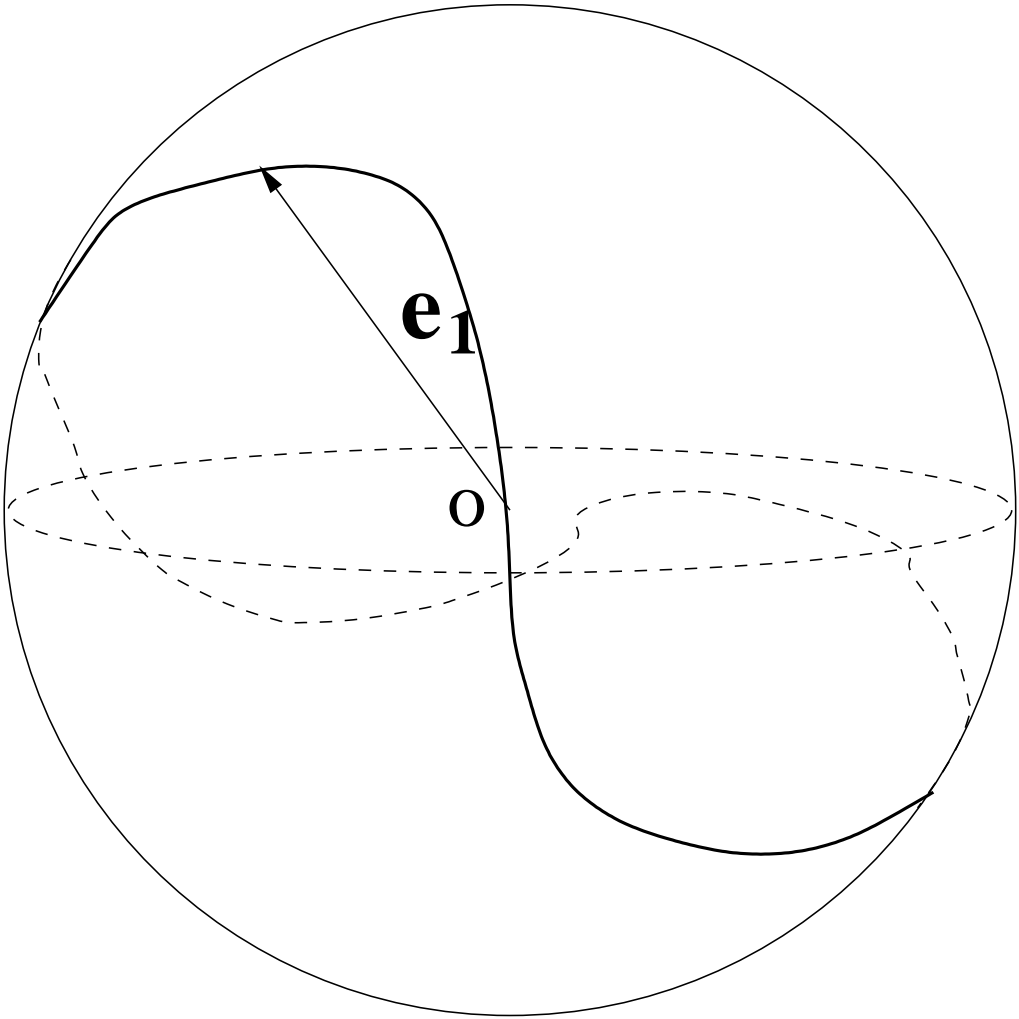
\begin{figure}
\caption{A closed curve on a unit sphere traced out by the tangent vector 
corresponding to a closed curve ${\bf R}(s)$ in 3-D.}
\end{figure}

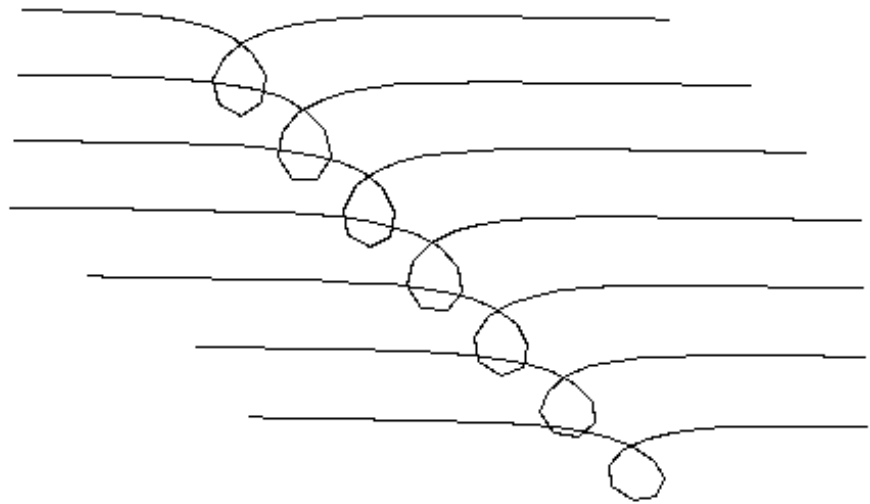
\begin{figure}
\caption{ A one-soliton filament vortex in a fluid,
corresponding to the one soliton solution 
$\psi=2\nu{\rm sech}(\nu(s-2\lambda t))\exp i(\lambda s +(\nu^2-\lambda^2)t)$.}
\end{figure}

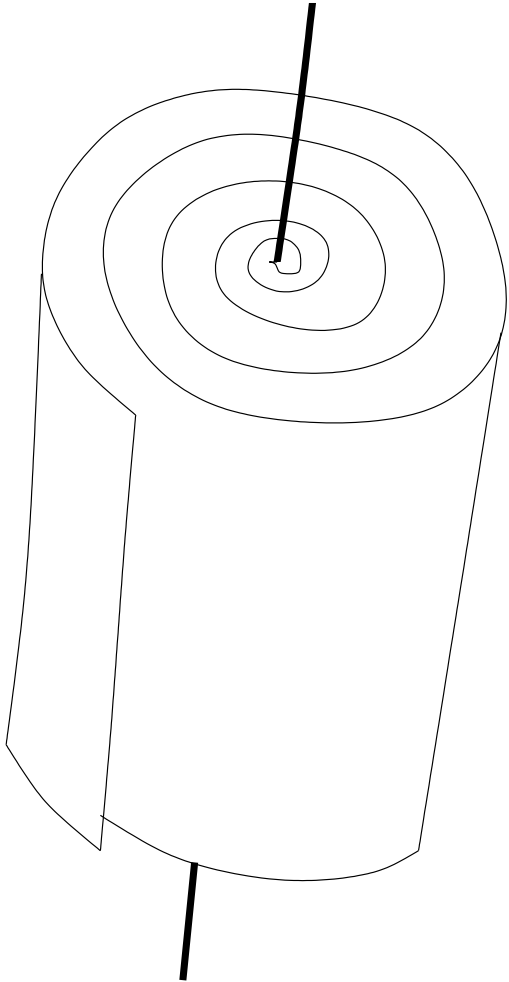
\begin{figure}
\caption{ A scroll wave. The scroll can be described as a spiral wave around
a line filament}
\end{figure}

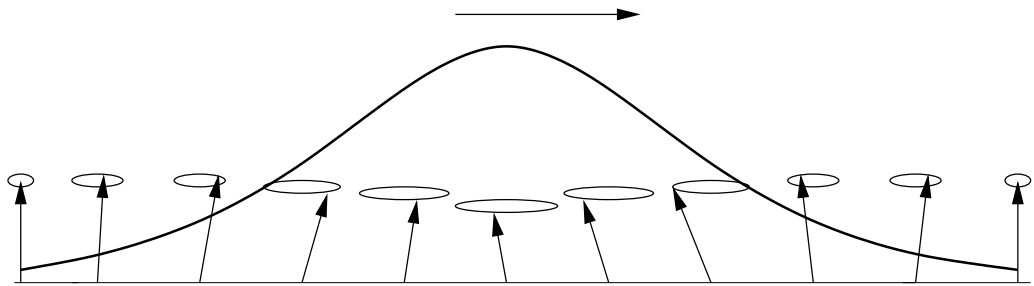
\begin{figure}
\caption{A propagating one soliton solution in a one dimensional
Heisenberg ferromagnetic spin chain. The spins rotate about the $\hat{z}$ axis.
The soliton profile describes the spin component in the $x-y$ plane 
$(1-S^3)$. }
\end{figure}

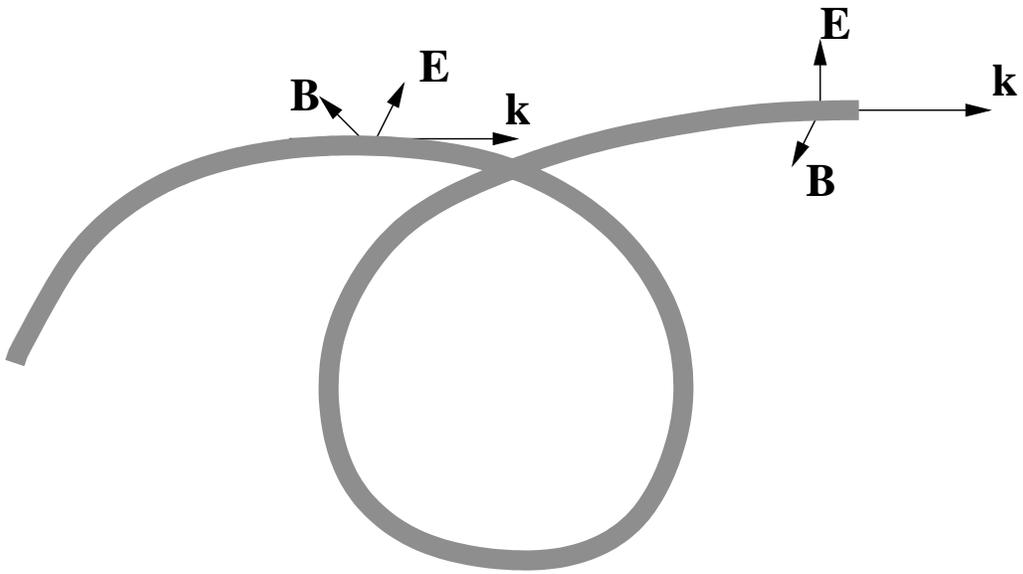
\begin{figure}
\caption{Propagation of polarized light along a curved optical 
fiber. The ${\bf E,B}$ fields acquire a geometric phase depending on the 
geometry of the fiber, given by $\int{\tau~ds}$.}
\end{figure}

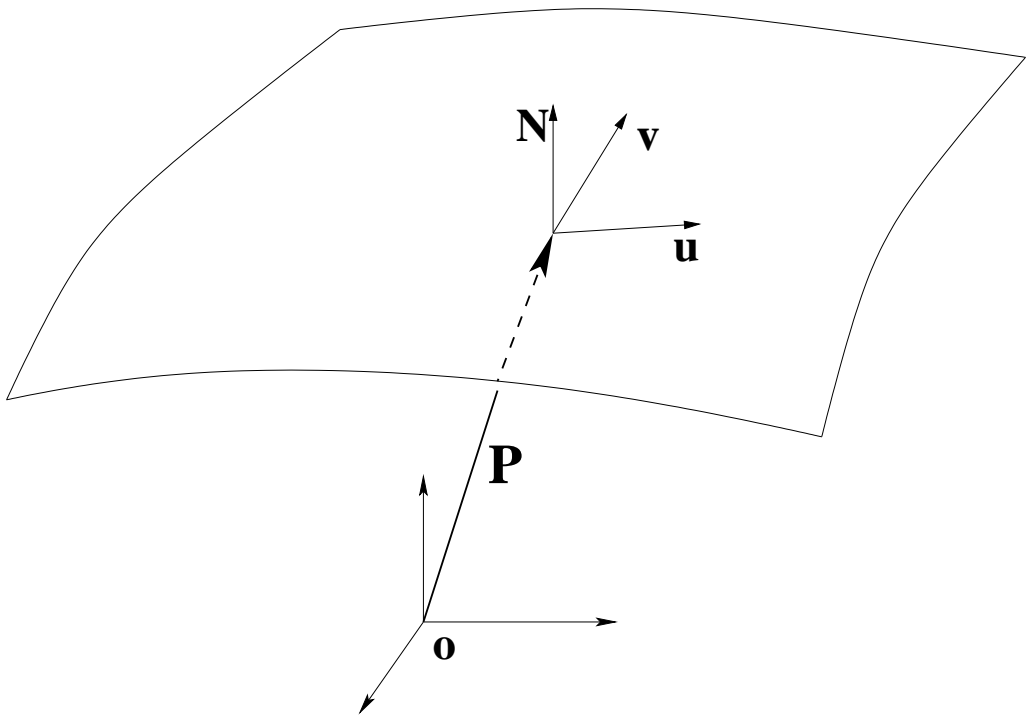
\begin{figure}
\caption{A surface is described by the position vector ${\bf P}(x,y)$. A pair
of tangent vectors ${\bf u, v}$ and a unit normal ${\bf N}$ can 
be defined at every point. }
\end{figure}

\end{document}